\documentclass{blois}
\pdfoutput=1 
\usepackage{jinstpub} 
\usepackage{lineno}

\usepackage{HGTD_TDR-defs}
\usepackage{multirow} 
\usepackage{rotating}
\usepackage{mathrsfs}
\usepackage{booktabs}
\usepackage{longtable}
\usepackage{placeins}
\usepackage{amssymb}
\usepackage{upgreek}
\usepackage{svg}
\usepackage{comment}
\usepackage{pdfpages}
\usepackage{siunitx}
\usepackage{lineno}
\usepackage{float}
\usepackage{lineno}
\usepackage{tikz}
\usepackage[percent]{overpic}
\usepackage{todonotes}
\usepackage{subcaption}

\title{\boldmath Measured gain suppression in FBK LGADs with different active thicknesses}
 
\author[1]{J.~Yang}
\author[4]{S.~Braun}
\author[1]{Q.~Buat}
\author[2]{J.~Ding}
\author[5]{M.~Harrison}
\author[1]{P.~Kammel}   
\author[2]{S.M.~Mazza}
\author[2]{F.~McKinney-Martinez}
\author[2]{A.~Molnar}
\author[3]{J.~Ott}
\author[2]{A.~Seiden}
\author[2]{B.~Schumm}
\author[4]{Y.~Zhao}
\author[6]{Y.~Zhang}
\author[6]{V.~Tishchenko}
\author[7]{A.~Bisht}
\author[7]{M.~Centis-Vignali}
\author[7]{G.~Paternoster }
\author[7]{M.~Boscardin}

\affiliation[1]{Center for Nuclear Physics and Astrophysics (CENPA), University of Washington, Seattle, WA 98195, USA}
\affiliation[2]{SCIPP, University of California Santa Cruz, 1156 High Street, Santa Cruz (CA), 950156, US}
\affiliation[3]{Department of Electrical and Computer Engineering, University of Hawaii at Manoa, 2540 Dole Street, Honolulu HI-96822, USA}
\affiliation[4]{Physics Department, Carleton University, 1125 Colonel By Drive, Ottawa, Ontario, K1S 5B6, Canada}
\affiliation[5]{Department of Physics and Astronomy, University of New Mexico, 210 Yale Blvd. NE, Albuquerque, NM 87106}
\affiliation[6]{Brookhaven National Laboratories, Upton (NY), U.S.}
\affiliation[7]{Fondazione Bruno Kessler}

\abstract{In recent years, the gain suppression mechanism has been studied for large localized charge deposits in Low-Gain Avalanche Detectors (LGADs). LGADs are a thin silicon detector with a highly doped gain layer that provides moderate internal signal amplification. Using the CENPA Tandem accelerator at the University of Washington, the response of LGADs with different thicknesses to MeV-range energy deposits from a proton beam were studied. Three LGAD prototypes of 50~$\mu$m, 100~$\mu$m, and 150~$\mu$m were characterized. The devices' gain was determined as a function of bias voltage, incidence beam angle, and proton energy.
This study was conducted in the scope of the PIONEER experiment, an experiment proposed at the Paul Scherrer Institute to perform high-precision measurements of rare pion decays.
LGADs are considered for the active target (ATAR), and energy linearity is an important property for particle ID capabilities.}


\notoc

\emailAdd{simazza@ucsc.edu}

\usepackage{sfmath}

\begin{document}
\maketitle

%\maketitle
%\flushbottom

\section{Introduction}
\label{sec:intro}

Low Gain Avalanche Detectors (LGADs) are thin silicon detectors with internal gain~\cite{bib:LGAD} capable of providing a time resolution as good as 17~ps~\cite{Zhao:2018qkg} for minimum-ionizing particles. LGADs are the chosen technology for near-future large-scale applications like the timing layer upgrade for HL\nobreakdash-LHC of the ATLAS~\cite{CERN-LHCC-2020-007} and CMS~\cite{CMS:2667167} experiments at CERN, the ePIC detector at the Electron\nobreakdash-Ion Collider at BNL~\cite{EIC} and the PIONEER experiment~\cite{Mazza:2021adt}.
In particular, this study was performed for the PIONEER collaboration since energy linearity is important for particle ID between positron, muon, and pion in the active target (ATAR)~\cite{Mazza:2021adt}.

This study is a continuation of a previous round of measurements done in 2023~\cite{Braun:2024sbi}; the saturation effect has also been studied by several other research groups~\cite{s22031080,Curr_s_2022}.
The gain saturation is triggered by the shielding of the electric field in the gain layer caused by the multiplication of the charge carriers in the bulk, and increases with the density of the deposited charge. Therefore, it depends on the total amount of initial deposited charge, the sensor gain, and the angle of incidence with respect to the charge drift direction towards the gain layer (the charge carrier drift is more distributed across the gain layer for an angled track).

\section{Experimental setup}
The Center for Experimental Nuclear Physics and Astrophysics (CENPA) at the University of Washington has a High Voltage Engineering Corporation Model FN tandem Van de Graaff accelerator~\cite{tandem} that is used to perform various accelerator-based experiments.
%In 1995, it was adapted to use an (optional) terminal ion source and a non-inclined tube 3, which enables the accelerator to produce high-intensity beams of hydrogen and helium isotopes at energies from 100 keV to 7.5 MeV.
Protons of 1.8~MeV, 3~MeV, 3.8~MeV, 4.5~MeV, 5.5~MeV, and 6.5~MeV momenta were used for this measurement campaign for devices that have an active thickness of \SI{50}{\micro\meter}, \SI{100}{\micro\meter}, and \SI{150}{\micro\meter}.
The momentum resolution of the proton beam is around 300~ppm, and the beam size is around 2~mm, which is smaller than the Rutherford back-scattering (RBS) target size and comparable with the sensor size.
%\todo{shall we compare this to the sensor size?}. 
%The calculated/simulated energy deposit of proton in 50~$\mu m$ Silicon substrate as a function of angle is shown in Fig.~\ref{fig:alpha_beta}, Left.

The tested LGAD sensors are mounted on a fast analog single channel electronic board (around 2~GHz bandwidth) designed at SCIPP~\cite{Padilla_2020} with an additional external box amplifier\footnote{GALI-52+ evaluation board}, the signal is digitized by the Sampic digitizer board~\cite{Breton:2020kva} with 11~bit, 1-10~Gs/s digitizer
%\todo{the latest one is https://www.sif.it/riviste/sif/ncc/econtents/2020/043/01/article/6 (and it is endorsed by the SAMPIC experts as a good ref)} \textcolor{red}{\bf ADD REFERENCE and some details}\todo{what details?} 
and a self-triggering capability.
The experimental setup is the same as in the previous article~\cite{Braun:2024sbi} using the Rutherford Backscattering (RBS) technique with 0.093~nm and 0.166~nm thick gold foils. 
The reference data for beta particles from a $^{90}$Sr source was taken in the SCIPP laboratories with identical sensors and readout boards; the energy of the $^{90}$Sr electron is in the minimum ionizing particle (MIP) range. The $^{90}$Sr system has been previously described in detail in~\cite{Zhao:2018qkg}. The CENPA test beam data and the $^{90}$Sr data were analyzed using the package described in the ``Data analysis'' section of the previous article~\cite{Braun:2024sbi}.

\section{Sensors tested}
\label{sec:devices}
Three single pad LGADs and three PINs (identical geometry but without gain layer) from FBK (Fondazione Bruno Kessler), were tested. The devices tested had \SI{50}{\micro\meter},  \SI{100}{\micro\meter}, and  \SI{150}{\micro\meter} of active thickness.
A list of the devices' characteristics is shown in Tab.~\ref{tab:LGADs}.
%A shallow and deep gain layer in the table means a gain layer with peaking doping within \SI{1}{\micro\meter} and over \SI{2}{\micro\meter} from the top surface of the sensor, respectively.
%The sensor with a deeper gain layer has an increased gain due to the effects discussed in literature~\cite{Padilla_2020}.
The table shows the breakdown voltage corresponding to the highest possible sensor gain. 
The energy deposited in the PIN detector depends on different beam energies and angles.
%is shown in Fig.~\ref{fig:stopping_sim}
%\todo{the current figure only shows simulated}. 
E.g., in the \SI{50}{\micro\meter} detector, At 1.8~MeV of energy, the proton always stops (stopping power 120~MeV/cm$^2$g, range in silicon around \SI{40}{\micro\meter}) and deposits about 80~fC of charge; at 3~MeV, the proton initially punches through and stops for an angle of approximately 55~degrees (stopping power 84.33~MeV/cm$^2$g, range in silicon around \SI{90}{\micro\meter}), depositing around 130~fC. 
In the \SI{100}{\micro\meter} detector, 3~MeV protons will always stop, 4.5~MeV protons will stop after 50~degrees, and 6.5~MeV protons will pass through even after 70~degrees.
In the \SI{150}{\micro\meter} detector, 5.5~M will stop after 50~degrees and 6.5~MeV will stop after 60~degrees.
%
%Proton stopping power and ranges were calculated with pstar~\cite{pstar}.
To provide a reference, the collected charge for a MIP is around 0.5~fC/1~fC/1.5~fC for the \SI{50}{\micro\meter}/\SI{100}{\micro\meter}/\SI{150}{\micro\meter} active thickness of the detector tested (Fig.~\ref{fig:sr90}, Left); therefore, the injected charge by the proton is around 160 and 260~MIP at 1.8~MeV and 3~MeV for a \SI{50}{\micro\meter} device, respectively.

\iffalse
\begin{figure}[h]
    \centering
    \includegraphics[width=0.32\textwidth]{CENPA_figures_2024/shifted_W1_1p8MeV.pdf}
    \includegraphics[width=0.32\textwidth]{CENPA_figures_2024/shifted_W1_3MeV.pdf}\\
    \includegraphics[width=0.32\textwidth]{CENPA_figures_2024/shifted_W8_3MeV.pdf}
    \includegraphics[width=0.32\textwidth]{CENPA_figures_2024/shifted_W8_4p5MeV.pdf}
    \includegraphics[width=0.32\textwidth]{CENPA_figures_2024/shifted_W8_6p5MeV.pdf}\\
    \includegraphics[width=0.32\textwidth]{CENPA_figures_2024/shifted_W14_5p5MeV.pdf}
    \includegraphics[width=0.32\textwidth]{CENPA_figures_2024/shifted_W14_6p5MeV.pdf}
    \caption{PIN (no gain) data and SRIM simulation of energy deposited by a proton with several initial energies in 50-100-150~$\mu m$ of Si as a function of incidence angle. The simulation and data do not perfectly match and are scaled to match the shape at a low angle.}
\label{fig:stopping_sim}
\end{figure}
%\todo[inline]{maybe we do not need all these plots or we can put some in appendix?}
\fi

\begin{table}[h]
    \centering
    \begin{tabular}{|l|c|c|c|c|c|}
        \hline
         Device & Producer & BV & Thickness & Gain layer & Gain range \\
         \hline
         W1 LGAD & FBK & \SI{250}{\volt} & \SI{50}{\micro\meter} & very shallow & 10-50\\
         W1 PIN & FBK & \SI{400}{\volt} & \SI{50}{\micro\meter} & no gain & 1\\
         \hline
         W8 LGAD & FBK & \SI{350}{\volt} & \SI{100}{\micro\meter} & very shallow & 5-30\\
         W8 PIN & FBK & \SI{500}{\volt} & \SI{100}{\micro\meter} & no gain & 1\\
         \hline
         W11 LGAD & FBK & \SI{600}{\volt} & \SI{150}{\micro\meter} & very shallow & 5-20\\
         W14 PIN & FBK & \SI{700}{\volt} & \SI{150}{\micro\meter} & no gain & 1\\
         \hline
    \end{tabular}
    \caption{List of tested FBK LGADs and PIN devices. All devices have a 2.5$\times$2.5~\si{\milli\meter\squared} single pad geometry. W11 and W14 have the same thickness but a difference in the gain layer. Therefore, PINs (no gain layer) from these two wafers are identical.}
    \label{tab:LGADs}
\end{table}

\section{$^{90}$Sr source Results}
\begin{figure}[h]
    \centering
    \includegraphics[width=0.45\textwidth]{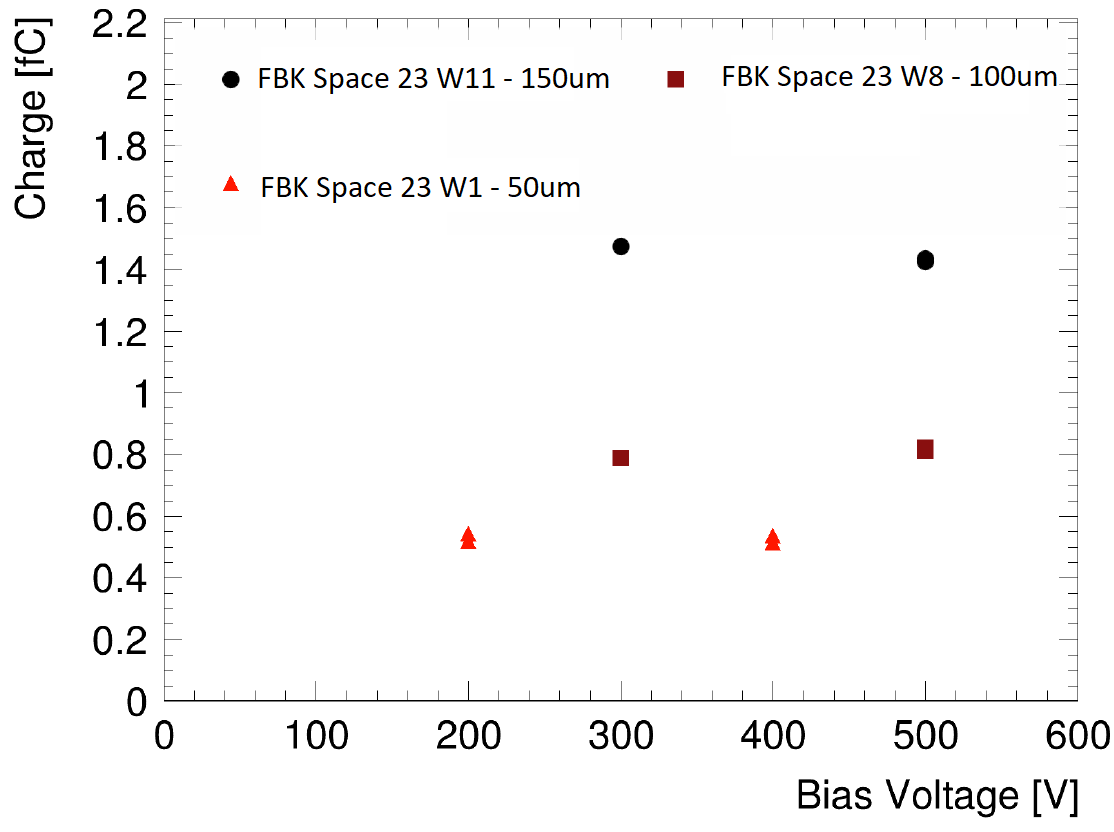}
    \includegraphics[width=0.45\textwidth]{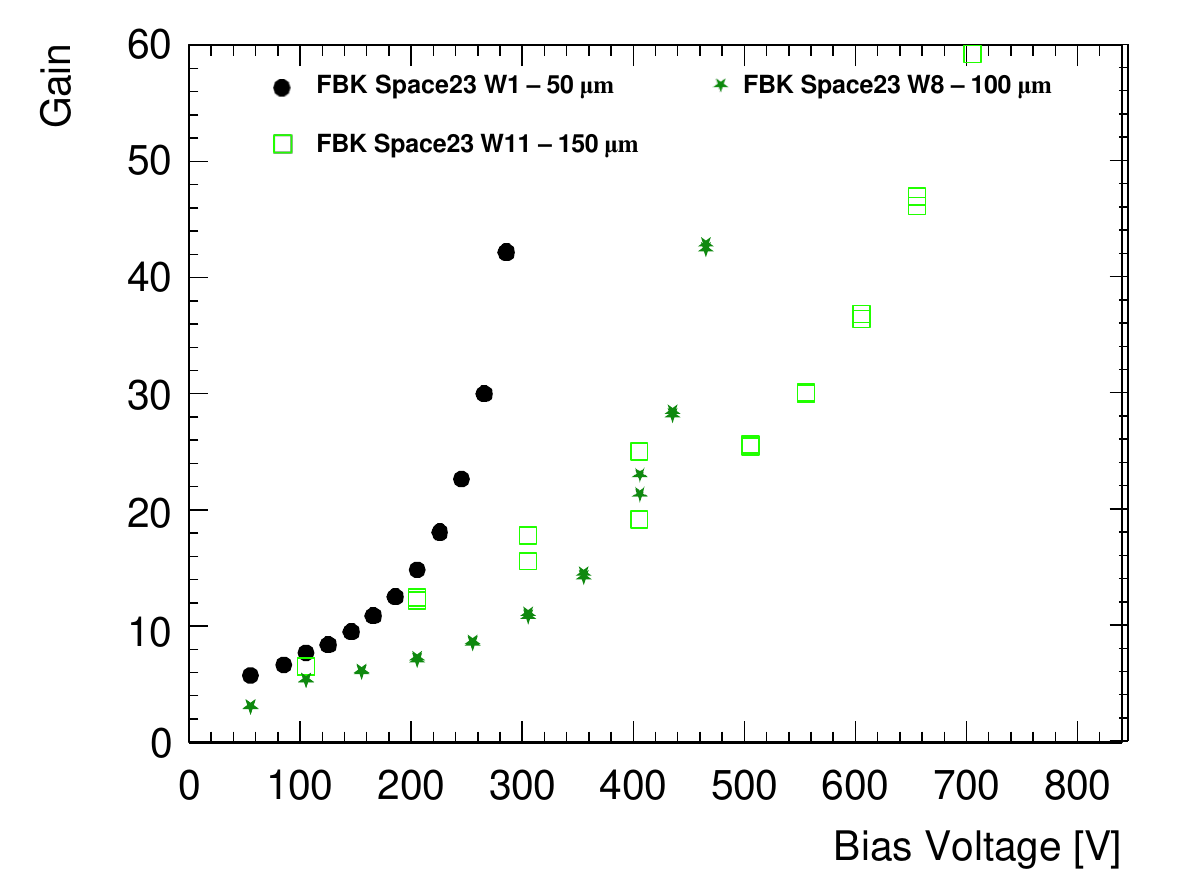}
    \caption{Left: Collected charge for Sr90 electrons as a function of bias voltage for FBK PIN detectors. Right: Gain for Sr90 electrons as a function of bias voltage for the FBK LGAD detectors. Data was taken with Sr90 source at SCIPP at 0 degrees incident angle.}
\label{fig:sr90}
\end{figure}

The sensors were tested in a laboratory with a $^{90}$Sr radioactive source in a coincidence system to ensure observation of MIP response. The setup is composed of two identical readout boards, one with the device-under-test (DUT) and one with a known trigger sensor, aligned with a $^{90}$Sr source. A full explanation of the whole setup can be found in Ref.~\cite{Zhao:2018qkg}. The collected charge vs bias voltage is shown in Fig.~\ref{fig:sr90}, left for PIN devices.  The charge values reflect the PIN thickness at \SI{50}{\micro\meter} and \SI{150}{\micro\meter} (around 0.1~fC per \SI{10}{\micro\meter}), but it's off for \SI{100}{\micro\meter}. The specification of the sensor thickness from FBK might not be completely correct for the \SI{100}{\micro\meter} device that seems to behave as being \SI{80}{\micro\meter} thick.
The gain (collected charge in the LGAD divided by the collected charge of a same-thickness PIN) vs bias voltage is shown in Fig.~\ref{fig:sr90}, Right.
Measurements were made only at a 0-degree (perpendicular) incident angle.

%\todo[inline]{Can you improve the legend? In particular on the right plot}

\section{Results}
\begin{figure}[]
    \centering
    \begin{subfigure}[b]{\textwidth}
    \centering
        \includegraphics[width=0.32\textwidth]{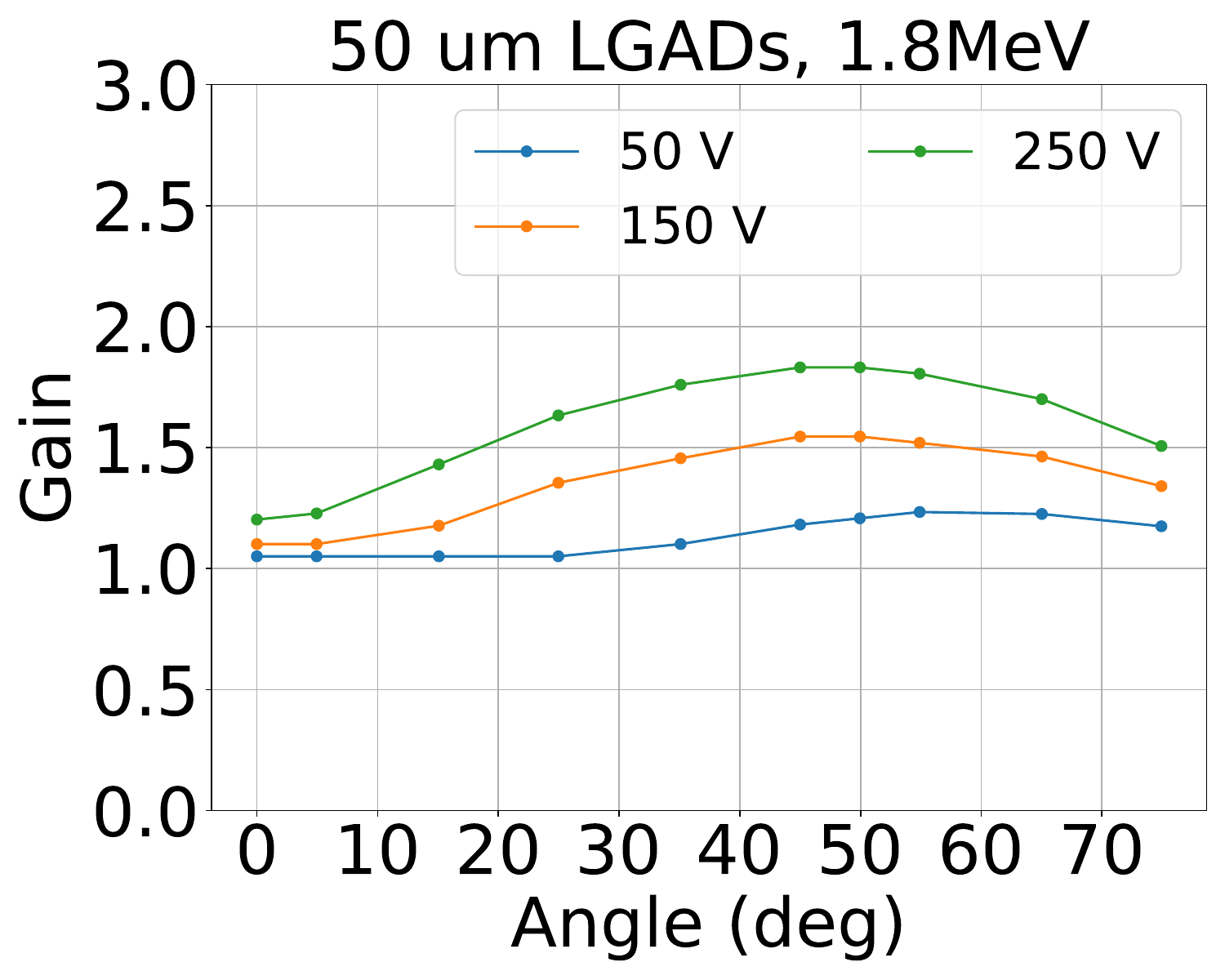}
        \includegraphics[width=0.32\textwidth]{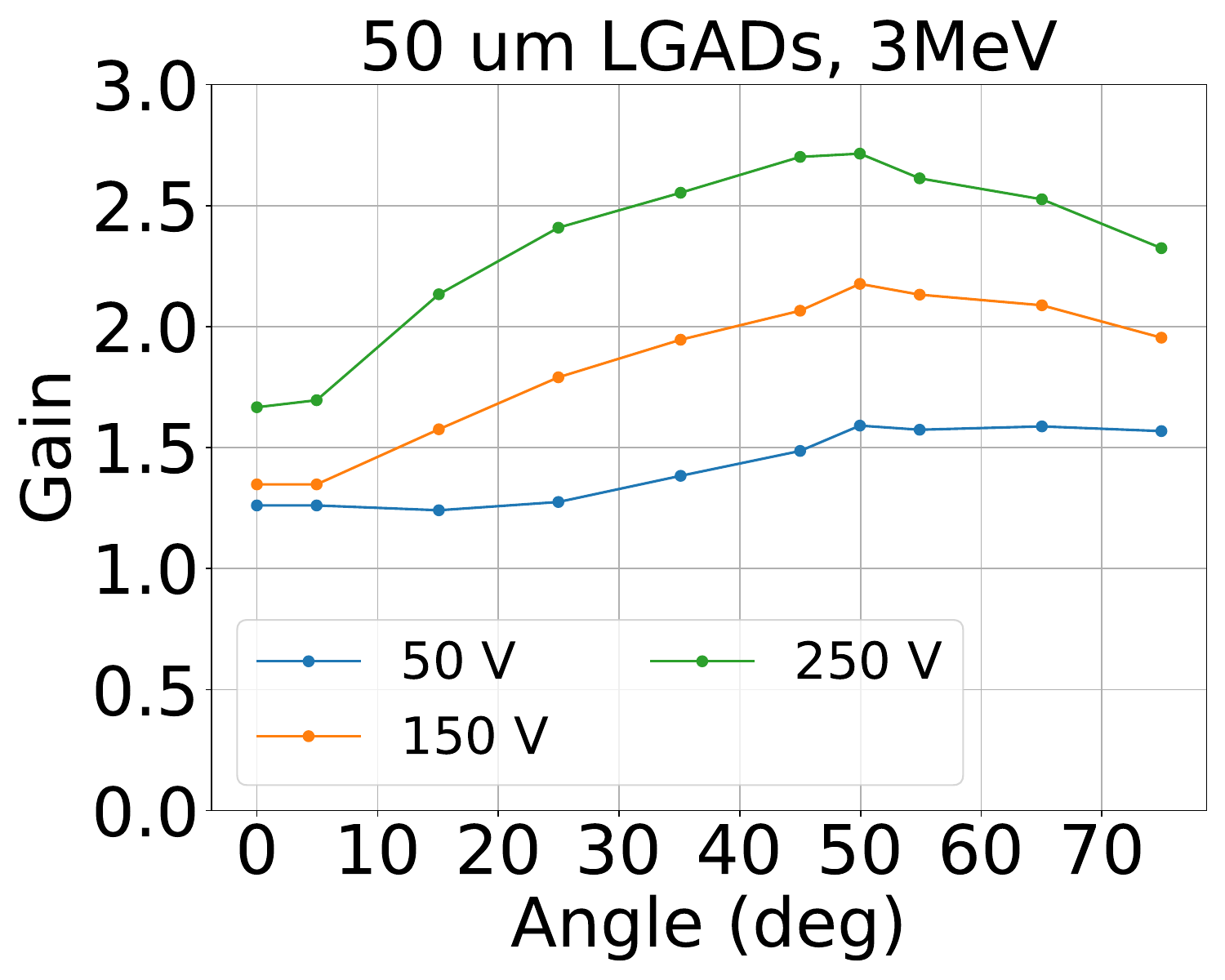}
        \caption{}
        \label{fig:HPK3p1_1p8MeV}
    \end{subfigure}
    \begin{subfigure}[b]{\textwidth}
    \centering
        \includegraphics[width=0.32\textwidth]{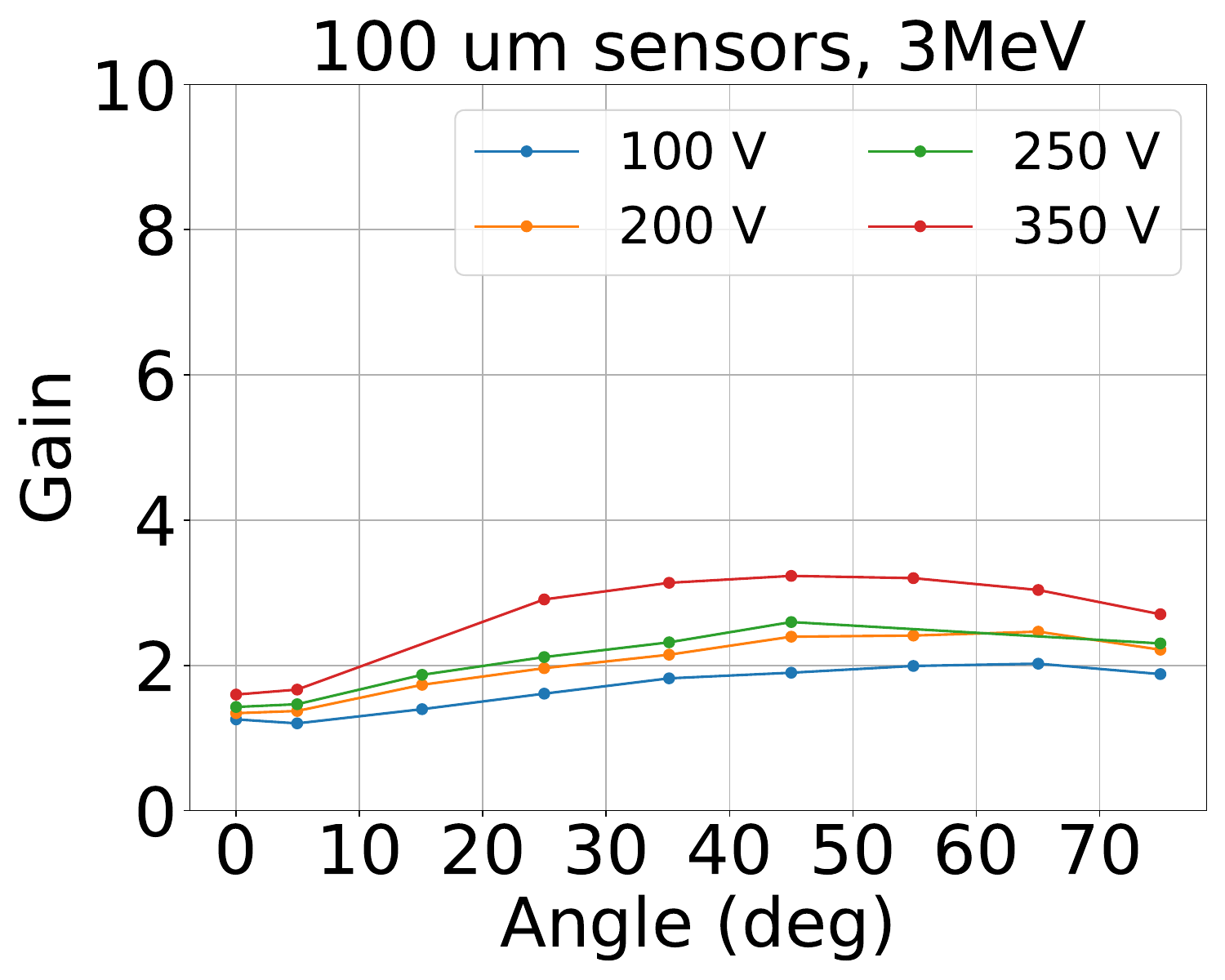}
        \includegraphics[width=0.32\textwidth]{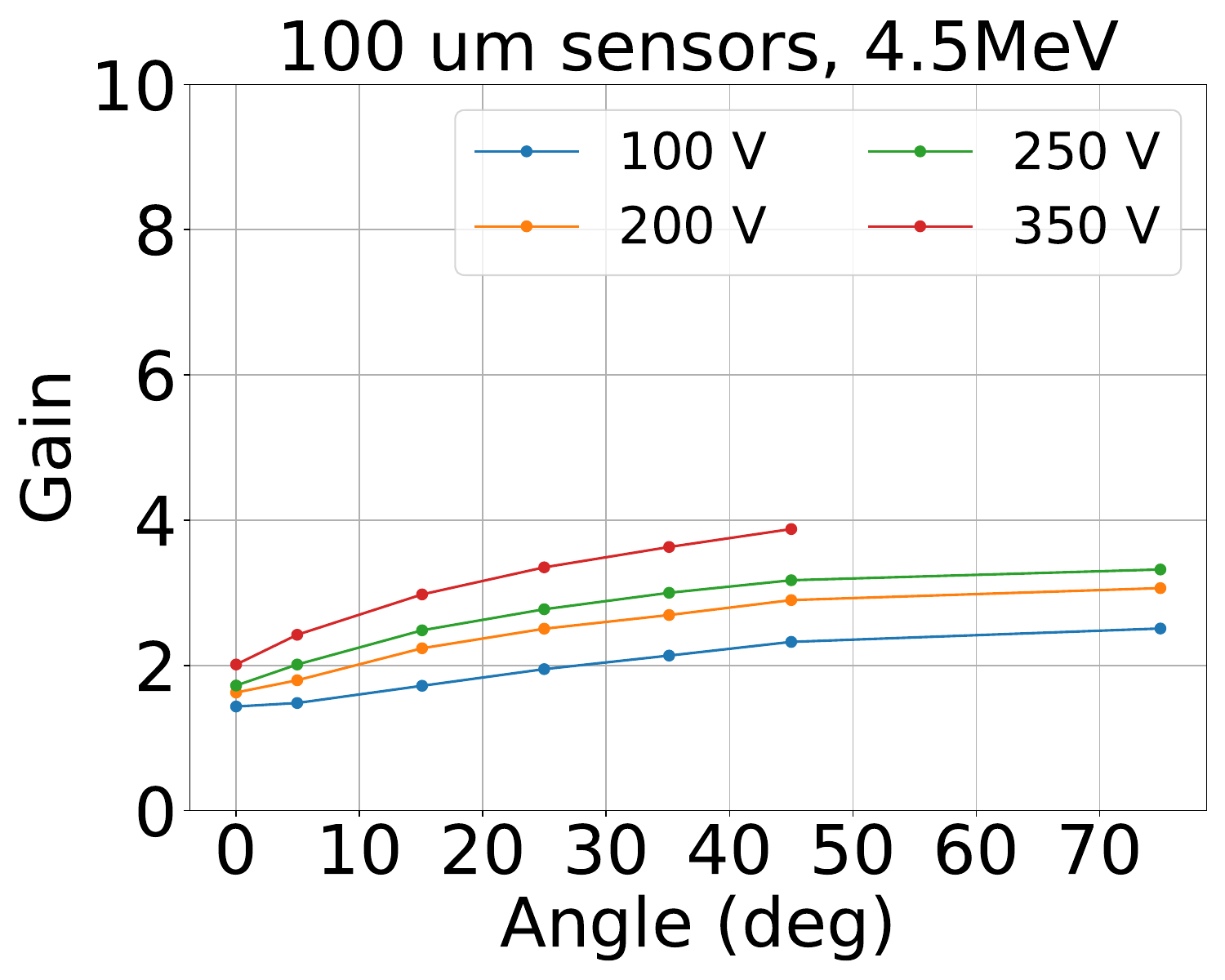}
        \includegraphics[width=0.32\textwidth]{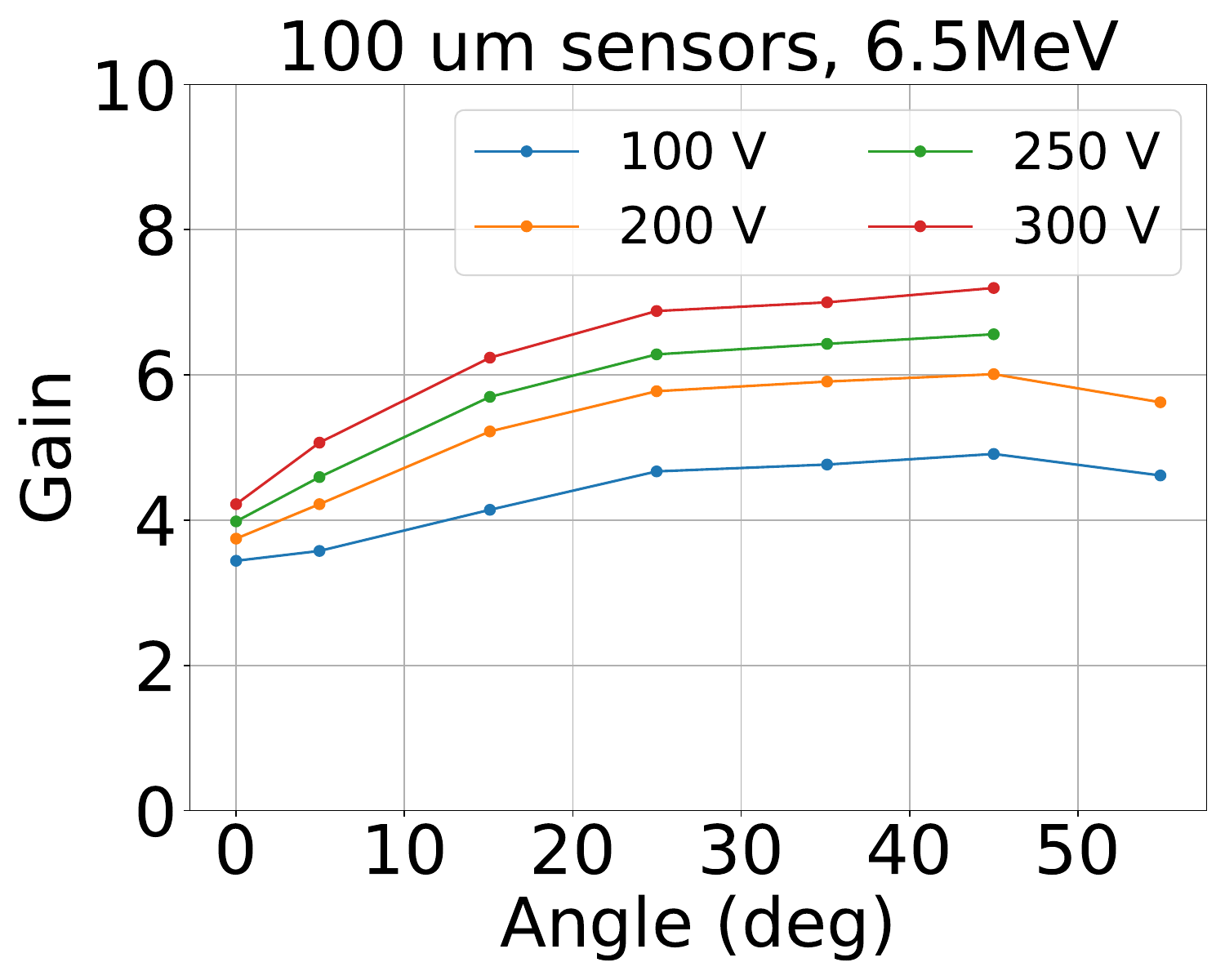}
        \caption{}
        \label{fig:HPK3p1_3MeV}
    \end{subfigure}
    \begin{subfigure}[b]{\textwidth}
    \centering
        \includegraphics[width=0.32\textwidth]{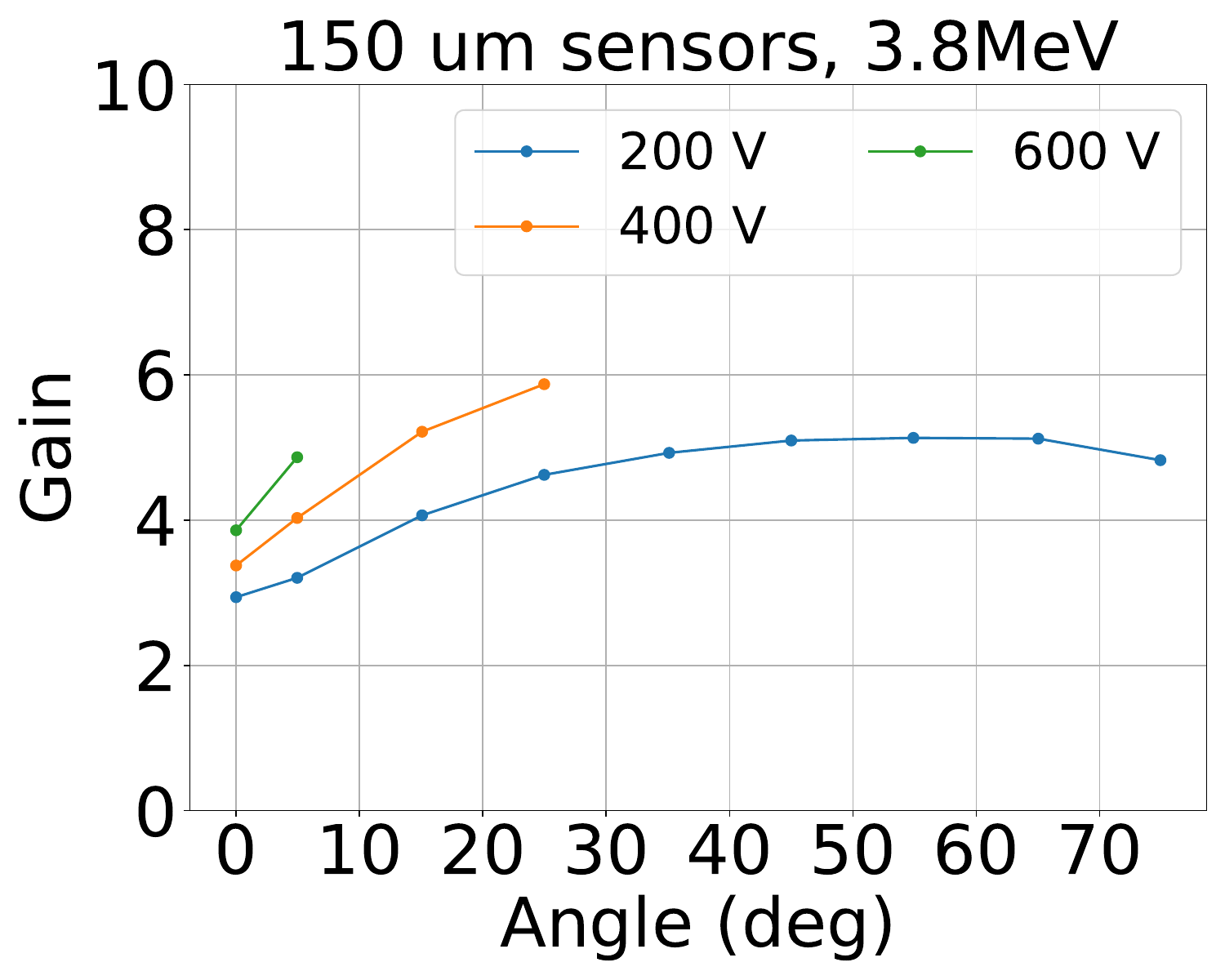}
        \includegraphics[width=0.32\textwidth]{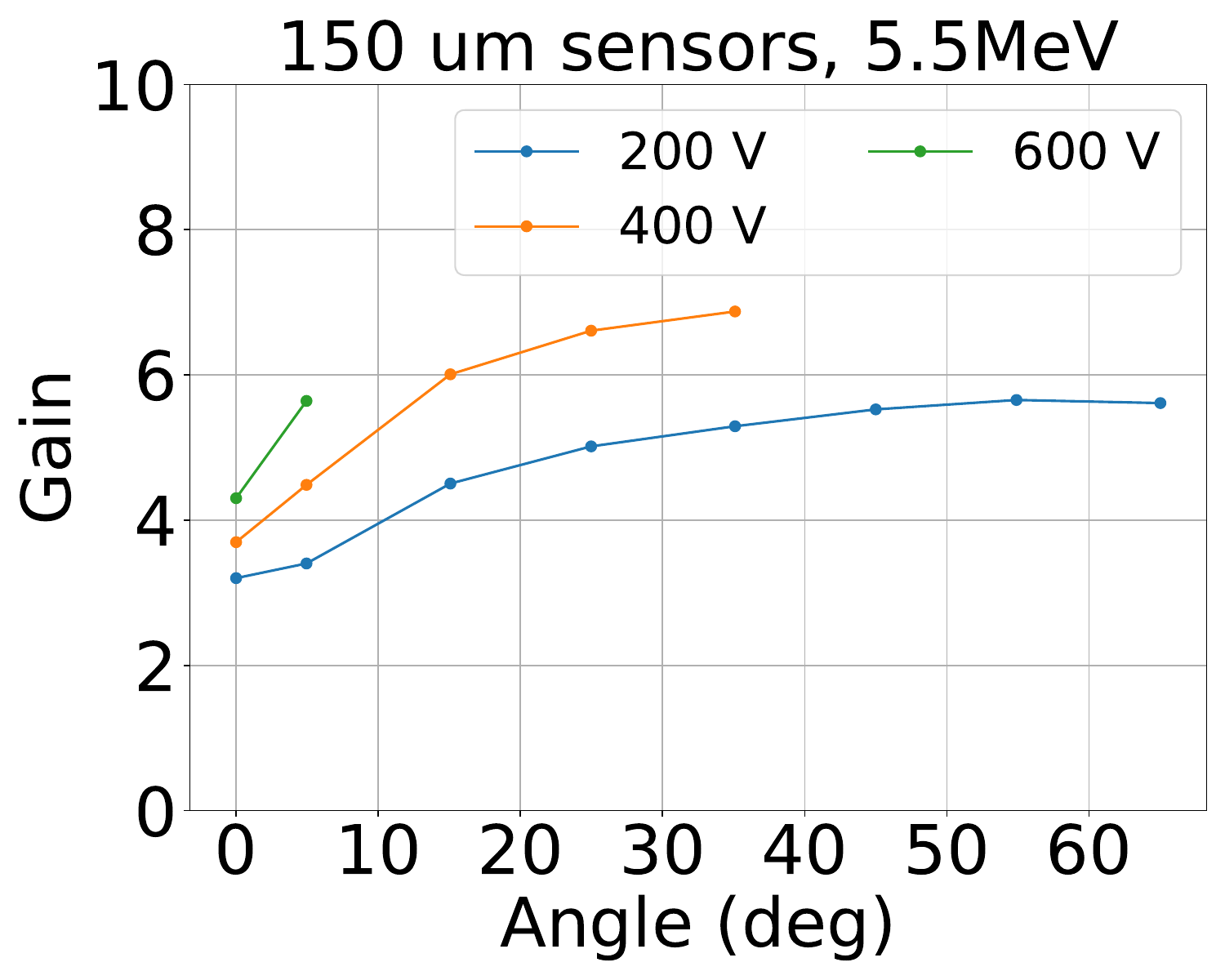}
        \caption{}
        \label{fig:HPK3p1_1p8MeV}
    \end{subfigure}
    \caption{Gain as a function of angle for different bias voltages for the three FBK LGADs at different energies. Particular runs were not included due to saturation in either the amplifier or digitizer for larger gains.} 
    \label{fig:gain1}
\end{figure}

The behavior of the devices was probed as a function of the proton incidence angle and applied bias voltage. 
The PIN sensors were tested for a limited set of voltages and showed almost no variation in the total charge deposition.
%, as seen in Fig.~\ref{fig:alpha_beta} (a). 
The gain as a function of angle for the \SI{50}{\micro\meter}, \SI{100}{\micro\meter}, and \SI{150}{\micro\meter} active thickness LGAD devices for several beam energies is shown in Fig.~\ref{fig:gain1} (a), (b), and (c), respectively.
The gain increases with the angle of incidence up to a certain angle; then, it decreases up to the largest angle tested (75 degrees).

The electrons drifting toward the gain layer are a projection of the entry track; for a zero-degree angle, the charge density on the gain layer is highly localized, while angled tracks produce a more dispersed and, therefore, reduced charge density throughout the gain layer, so the suppression mechanism is less pronounced.
The gain decrease at large angles can be explained by the competing effect of the charge deposition's proximity to the gain layer. In Ref.~\cite{Mazza:2023col, Molnar:2025txf}, it was shown that the depth of X-ray absorption influences the sensor response such that the gain is effectively less for charge depositions closer to the gain layer. The charge lateral diffusion during drift towards the gain layer is the most likely explanation for this effect.
The proton energy deposition profile has the properties of a Bragg peak, depositing a large amount of energy at the stopping point. 
At a large angle, the proton deposits a larger amount of charge closer to the gain layer, given that the track length is always the same when the ion stops inside the active layer.

\section{Conclusions}

The response to highly ionizing particles of several types of FBK LGADs and PINs was studied at the CENPA tandem accelerator with protons of several different energies.
The gain in the LGADs was calculated as a function of proton energy, applied bias voltage, and angle of incidence and compared to the response of the same-size PIN detector.
All LGAD devices show substantial gain suppression with respect to the gain for MIPs measured in laboratory, which is more significant for higher applied bias voltage.
The suppression changes with the angle of incidence of the beam: it increases up to 40 to 60 degrees of incidence, but then it is again reduced because of the large deposition closer to the gain layer.

Comparing the results of this paper with the previous results for HPK sensors~\cite{Braun:2024sbi}, the gain saturation effect is significantly more prominent, especially for the direct comparison with the \SI{50}{\micro\meter} active layer device with a similar breakdown voltage. 
A possible explanation of the observed behavior is the difference in doping profile between HPK and FBK sensors; the latter have a very shallow gain layer\footnote{Detailed doping profiles are protected information of HPK and FBK, so they cannot be disclosed}. 
A shallow gain layer has a very compact high-field region, which leads to a higher charge density after the multiplication starts, causing a higher suppression effect.

A series of detectors fabricated for ATLAS and CMS timing layers from HPK and FBK with different doping concentrations and profiles will be considered for the next data-taking campaign. This focused study will probe the gain suppression for different gain layer configurations; the final goal is to fabricate a device that is as linear as possible on a large range of deposited energies for the PIONEER experiment.

%\textit{summary of results, some conclusions}

\section{Acknowledgments}
This work was supported by the United States Department of Energy grants DE-SC0010107, DE-FG02-04ER41286, and DE‐FG02‐97ER41020. We thank the CENPA Technical and Accelerator staff for their assistance in this measurement.
%C.~Lansdell received undergraduate researcher support through the NSF REU Program at UW, NSF Grant No 2243362.
%\textcolor{red}{other funding agency acknowledgements? Thanks to CENPA staff?}

\small
\bibliography{bib/TechnicalProposal,bib/hpk_fbk_paper,bib/HGTD_TDR,bib/SHIN,bib/nizam,bib/others}

\end{document}